\documentclass[apj]{emulateapj}   % somehow this puts it on one page
\usepackage{graphicx}
\usepackage{multirow}

\usepackage{amsmath}
\usepackage{natbib}
\usepackage{hyperref}
\usepackage{physics}
\usepackage[usenames,dvipsnames]{color}  %SdM: allows fancy color names
\usepackage{lineno}
\usepackage[normalem]{ulem}

\def\msun{{ ~M}_{\odot}}
\def\rsun{{ ~R}_{\odot}}

\def\kms{{\rm ~km} {\rm ~s}^{-1}}
\def\mpy{{\rm ~M}_{\odot} {\rm ~yr}^{-1}}

\definecolor{Orange-red}{rgb}{1.0, 0.27, 0.0}

\begin{document}

\title{The implications of high BH spins on the origin of BH-BH mergers}

\author{
   A. Olejak\altaffilmark{1}, K. Belczynski\altaffilmark{1} 
}

\affil{
   $^{1}$ Nicolaus Copernicus Astronomical Center, Polish Academy of Sciences,
          ul. Bartycka 18, 00-716 Warsaw, Poland, 
          (aleksandra.olejak@wp.pl,chrisbelczynski@gmail.com)
}

\begin{abstract}

The LIGO/Virgo collaboration has reported $50$ black hole---black hole (BH-BH) mergers and $8$ candidates recovered from digging deeper into the detectors noise. The majority of these 
mergers have low effective spins pointing toward low BH spins and efficient angular 
momentum transport (AM) in massive stars as proposed by several models (e.g., the Tayler-Spruit 
dynamo). However, out of these $58$ mergers, $7$ are consistent with having
high effective spin parameter ($\chi_{\rm eff}>0.3$). Additionally, $2$  
events seem to have high effective spins sourced from the spin of the primary (more massive) BH. 
These particular observations could be used to discriminate between the isolated binary and dynamical formation channels. It might seem that high BH spins point to a dynamical origin if AM in stars is efficient and forms low-spinning BHs. In such a case dynamical formation is required to produce second and third generations of BH-BH mergers with typically high-spinning BHs.
Here we show, however, that isolated binary BH-BH formation naturally reproduces such highly spinning BHs. Our models start with efficient AM in massive stars 
that is needed to reproduce the majority of BH-BH mergers with low effective spins. Later, some of the binaries are subject to a tidal spin-up allowing the formation of a moderate fraction 
($\sim 10\%$) of BH-BH mergers with high effective spins ($\chi_{\rm eff}\gtrsim0.4-0.5$). 
In addition, isolated--binary evolution can produce a small fraction of BH-BH mergers with almost maximally spinning primary BHs.
Therefore, the formation scenario of these atypical BH-BH mergers remains to be found.

\end{abstract}


\section{Introduction}
\label{sec.intro}

The LIGO/Virgo collaboration has announced detection of gravitational waves from $\sim 50$ double black hole (BH-BH) mergers~\citep{LIGO2019a,LIGO2019b,2020ApJ...891L..27F,Abbott2021a}. Additional $8$ BH-BH merger candidates have been recently reported ~\citep{Abbott2021b}. The majority of all these events have low effective spins parameters: $\chi_{\rm eff}= \frac{m_1 a_{\rm 1} \cos \theta_1+m_2 a_{\rm 2} \cos \theta_2}{m_1 + m_2} \approx 0,$
where $m_{i}$ are BH masses, $a_{\rm i}=cJ_{i}/Gm_{i}^2$ are dimensionless BH spin magnitudes 
($J_{i}$ being the BH angular momentum (AM), $c$ the speed of light, $G$ the gravitational constant), and 
$\theta_{i}$ are angles between the individual BH spins and the system orbital AM. 

However, among the dectections there are also several BH-BH mergers which are characterized by higher (non-zero) positive effective spins. In Table 1 we list 
the parameters of the five BH-BH mergers with highest effective spins reported by \cite{Abbott2021a} with additional two high effective spin systems reported by \cite{Abbott2021b}.

\begin{table}
\caption{BH-BH mergers with high effective spins}
\begin{tabular}{clcccc}
\hline\hline
No. & Name$^{a}$ & $\chi_{\rm eff}$ & $m_1$ & $m_2$ & $a_1$ \\
\hline\hline
1 & GW190517 & $0.52^{+0.19}_{-0.19}$ & $37.4^{+11.7}_{-7.6}$  & $25.3^{+7.0}_{-7.3}$  & -- \\
2 & GW170729         & $0.37^{+0.21}_{-0.25}$ & $50.2^{+16.2}_{-10.2}$ & $34.0^{+9.1}_{-10.1}$  & -- \\
3 & GW190620 & $0.33^{+0.22}_{-0.25}$ & $57.1^{+16.0}_{-12.7}$ & $35.5^{+12.2}_{-12.3}$  & -- \\
4 & GW190519 & $0.31^{+0.20}_{-0.22}$ & $66.0^{+10.7}_{-12.0}$ & $40.5^{+11.0}_{-11.1}$  & -- \\
5 & GW190706 & $0.28^{+0.26}_{-0.29}$ & $67.0^{+14.6}_{-13.3}$ & $38.2^{+14.6}_{-13.3}$  & -- \\
                      &&&&& \\
6 & GW190403 & $0.70^{+0.15}_{-0.27}$ & $88.0^{+28.2}_{-32.9}$ & $22.1^{+23.8}_{-9.0}$  & $0.92^{+0.07}_{-0.22}$ \\ 
7 & GW190805 & $0.35^{+0.30}_{-0.36}$ & $48.2^{+17.5}_{-12.5}$ & $32.0^{+13.4}_{-11.4}$ & $0.74^{+0.22}_{-0.60}$ \\ 
\hline
\hline
\end{tabular}
\\$^{a}$: Names are abbreviated. We include candidate detections as full
astrophysical events. Parameters of first $5$ events are from original LIGO/Virgo analysis \citep{Abbott2021a}, while the remaining $2$ are from deeper search into the detectors noise \citep{Abbott2021b}.\\
\label{tab.obs}
\end{table}

The formation of close BH-BH systems is an open issue with several formation channels proposed and 
discussed in the context of the LIGO/Virgo mergers. The major formation scenarios include the isolated binary evolution ~\citep{Bond1984b,Tutukov1993,Lipunov1997,Voss2003,Belczynski2010a,Dominik2012,Kinugawa2014,Hartwig2016,deMink2016,Mandel2016a,Marchant2016,Spera2016,Belczynski2016b,Eldridge2016,Woosley2016,Heuvel2017,Stevenson2017,Kruckow2018,Hainich2018,Marchant2018,Spera2019,Neijssel:2019,Buisson2020,Bavera2020,Bavera2021,Qin2021}, the dense stellar system dynamical channel~\citep{PortegiesZwart2000,Miller2002a,Miller2002b,PortegiesZwart2004,Gultekin2004,Gultekin2006,OLeary2007,Sadowski2008,Downing2010,Antonini2012,Benacquista2013,Mennekens2014,Bae2014,Chatterjee2016,Mapelli2016,Hurley2016,Rodriguez2016a,VanLandingham2016,Askar2017,ArcaSedda2017,Samsing2018,Morawski2018,Banerjee2018,DiCarlo2019,Zevin2019,Rodriguez2018a,Perna2019,Kremer2020a}, isolated multiple (triple, quadruple) systems ~\citep{Antonini2017b,Silsbee2017,Arca-Sedda2018,LiuLai2018,
Fragione19}, mergers of binaries in galactic nuclei \citep{Antonini2012b,Hamers2018,Hoang2018,Fragione2019b} and primordial BH formation~\citep{Sasaki2016,Green2017,Clesse2017,Carr2018,DeLuca2020}. 

BH spins and their orientations can play an important role in distinguishing between various BH-BH 
formation models. If the BH spins are not small, then their orientation may possibly distinguish between 
a binary evolution origin (predominantly aligned spins) and dynamical formation channels (more or less 
isotropic distribution of spin orientations). If the BHs formed out of stars have small spins~\citep{
Spruit2002,Zaldarriaga2017,Hotokezaka2017,Fuller2019a,Qin2019,Olejak2020b,Bavera2020,Belczynski2020b} 
then BH-BH mergers with high effective spins may challenge their isolated evolution origin. In dense stellar clusters, BHs may merge several times easily producing BHs with high spins and making a dynamical channel a prime site for such events \citep{Gerosa&Berti2017,Fishbach2017}. However, the assumption about the BH natal spin (and the AM transport efficiency) also plays a role in the effective spin distribution for the dynamical channel \citep{Benerjee2021}.

In this study we show that the current understanding of stellar/binary astrophysics \citep{Belczynski2021} and the degeneracy between the different formation channels do not allow for such a simple test of the origin of the LIGO/Virgo BH-BH mergers. To demonstrate this we show that although the isolated binary evolution channel produces mostly BH-BH mergers with low effective spins, a small but significant fraction of mergers is expected to have moderate or even high effective spins. Despite the assumption that stars slow down their rotation due to efficient AM transport, we find that tidal interactions are capable of spinning up some stars allowing formation of rapidly spinning BHs ~\citep{Detmers2008,
Kushnir2017,Qin2018}.

\section{Method}
\label{sec.calc}

We use the population synthesis code {\tt StarTrack}~\citep{Belczynski2002,Belczynski2008a} with a model of star formation rates and metallicity distribution based on \cite{Madau2014} described in \cite{Belczynski2020b}.
We employ the delayed core-collapse supernova (SN) engine for neutron star/BH mass calculation~\citep{Fryer2012}, 
with weak mass loss from pulsation pair instability supernovae~\citep{Belczynski2016c}. 
BH natal kicks are calculated from a Maxwellian distribution with $\sigma=265\kms$ and decreased by 
fallback during core-collapse; this makes a significant fraction of BHs form without a natal kick~\citep{Mirabel2003}. 
We assume our standard wind losses for massive O/B stars~\citep{Vink2001} and LBV winds \citep[specific prescriptions for these winds are listed in Sec.~2.2 of][]{Belczynski2010b}. BH natal spins are 
calculated under the assumption that AM in massive  stars is transported by the 
Tayler-Spruit magnetic dynamo~\citep{Spruit2002} as adopted in the MESA stellar evolutionary code
~\citep{Paxton2015}. Such BH natal spins take values in the range $a \in 0.05-0.15$  \cite[see][]{Belczynski2020b}. Note that the modified classic Tayler-Spruit dynamo with a different non-linear saturation mechanism of the Tayler instability \citep{Fuller2019a, Fuller2019b} causes larger magnetic field amplitudes, more efficient AM transport and even lower final natal spins $(a \sim 0.01)$.
BH spin may be increased if the immediate BH progenitors (Wolf-Rayet: WR) stars in close binaries are 
subject to tidal spin-up. In our calculations for BH-WR, WR-BH and WR-WR binary systems with orbital periods in the range 
$P_{\rm orb}=0.1-1.3$ d the BH natal spin magnitude is fit from WR star spun-up MESA models (see eq.15 
of \cite{Belczynski2020b}), while for systems with $P_{\rm orb}<0.1$d the BH spin is taken to be equal to $a=1$. 
BH spins may also be increased by accretion in binary systems. We treat accretion onto a compact object 
during Roche lobe overflow (RLOF) and from stellar winds using the analytic approximations presented 
in \cite{King2001} and \cite{Mondal2020}. In the adopted approach the accumulation of matter on a BH is very inefficient so accretion does not noticeably affect the final BH spin. However note that, e.g. \cite{vonSon2020} or \cite{Bavera2021} tested different super Eddington accretion prescriptions finding that some BHs may be significantly spun-up by accretion.

For common the envelope (CE) evolution we assume a $100\%$ ($\alpha_{\rm CE}=1$) orbital energy transfer for CE ejection 
and we adopt $5\%$ Bondi accretion rate onto the BHs during CE~\citep{Ricker2008,MacLeod2015a,MacLeod2017a}. 
During the stable RLOF (whether it is a thermal- or nuclear-timescale mass transfer: TTMT/NTMT) we 
adopt the following input physics. If an accretor is a compact object (neutron star or BH) we allow for 
super-Eddington accretion with excess transferred mass lost with an AM specific 
to the accretor~\citep{Mondal2020}. In all other cases, we allow a fraction of the transferred mass 
of $f_{\rm a}=0.5$ to be lost from the binary with a specific AM of the binary orbit
$j_{\rm loss}=1.0$ (expressed in units of $2 \pi A^2 / P_{\rm orb}$, $A$ being an orbital 
separation; see eq. 33 of \cite{Belczynski2008a}). 

RLOF stability is an important issue in the context of BH-BH system formation in the framework of 
the isolated binary evolution ~\citep{Neijssel:2019,Olejak2021a,Gallegos-Garcia:2021hti,Belczynski2021}. In the standard {\tt StarTrack} 
evolution we impose rather liberal limits for CE (dynamical-timescale RLOF) to develop (see
~\cite{Belczynski2008a}: binaries with a donor star more massive than $2-3$ times the mass of the accretor 
are subject to CE. In this model (for simplicity tagged here as CE model) the vast majority of BH-BH 
mergers form through CE evolution, although we find some cases ($\lesssim 1\%$) of BH-BH merger 
formation without any CE event. In the alternative model \citep[non-CE model, detailed description in][]{Olejak2021a} we allow CE to be suppressed 
for some systems even with mass ratio as high as $6-8$~\citep{Pavlovskii2017}. In this model the
majority of the BH-BH mergers form without any CE event (the orbital decrease is obtained through angular 
momentum loss during stable RLOF), although some ($< 10\%$) BH-BH mergers form with the
assistance of CE. 

For each model we calculate the evolution of $64$ million massive, Population I/II binary systems. We 
use the star formation history and chemical evolution of the Universe to obtain the BH-BH merger properties 
within an approximate reach of LIGO/Virgo (redshift $z<1$). We use the same method as described 
in~\cite{Belczynski2020b}.

\section{Results}
\label{sec.res}

\begin{figure}
\hspace*{-0.4cm}
\includegraphics[width=0.5\textwidth]{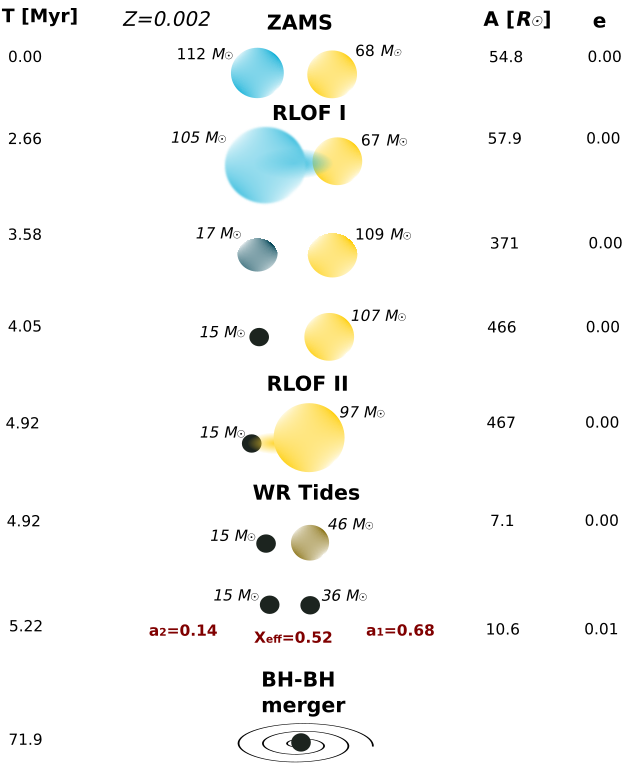}
\caption{
Typical example of non-CE evolutionary scenario leading to the formation of BH-BH merger with tidally 
spun-up primary: $a_1=0.68$ and $\chi_{\rm eff}=0.52$. Binary system goes through two phases of 
RLOF with episodes of nuclear and thermal timescale mass transfer. RLOF I ends with the system 
mass ratio reversal. After RLOF II the system orbital separation significantly decreases 
and WR star is a subject to tidal spin-up by a BH. Soon thereafter the close BH-BH system is 
formed with a short merger time of $\sim 67$ Myr (see Sec.~\ref{sec.res}).
}
\label{fig.evol1}
\end{figure}

Figure~\ref{fig.evol1} shows a typical example of binary system evolution without a CE phase leading 
to the formation a BH-BH merger with a tidally spun-up primary BH~\citep[restricted RLOF stability 
criteria;][]{Olejak2021a}. The rather unequal-mass massive stellar system ($112\msun$ and $68\msun$) 
with a metallicity of $Z=0.002$ goes through two RLOF events. The RLOF I is initiated 
by the more massive star; first by an NTMT when the donor is still on the main-sequence and then through a TTMT when the donor 
evolves off main-sequence. After the RLOF I, the system mass ratio is reversed: the initially more massive 
star lost over $80\%$ of its mass while the companion gained $\sim 40\msun$. Next, the initially 
more massive star ends its evolution directly collapsing to the less--massive (secondary) BH with a mass of $m_2=15\msun$ 
and spin $a_2=0.14$. When the companion star expands, it initiates a second stable RLOF. At the 
onset of RLOF II the system has highly unequal masses: the donor is almost $6.5$ times more massive 
than the BH. The thermal timescale for a donor with mass $M_{\rm don}\approx 97 \msun$, radius $R_{\rm don}\approx 300 \rsun$ and luminosity $L_{\rm don}\approx 3 \times 10^6 L_{\odot}$ (parameters at the RLOF II onset \footnote{Such parameter values are inline with other predictions for massive stars e.g. using Geneva stellar
evolution code \citep{Yusof}.} ) calculated with the formula by \citet{Kalogera:1995yc}, is
$ \tau_{\rm th} \approx 330$ $\rm{ yr}.$ 
It corresponds to a very high mass transfer rate $\dot{M}= M_{\rm don}/\tau_{\rm th} \approx 0.3\mpy$ which does not allow the BH to accrete much
mass (despite the fact that we allow for super-Eddington accretion). Half of the donors mass
is lost from the binary with the specific AM of the BH (as the matter was transferred to the 
vicinity of the BH accretor).
This has a huge effect on the orbital separation which decreases from $A=467\rsun$ 
to only $A=7.1\rsun$. After RLOF II the binary consists of a BH and a WR star that are close enough to allow 
for the tidal spin-up of the WR star. Finally, the WR star directly collapses to the more massive (primary) BH with 
a mass $m_1=36\msun$ and spin $a_1=0.68$. The BH-BH system merges after $\sim 67$ Myr.

\begin{figure}
\hspace*{-0.4cm}
\includegraphics[width=0.5\textwidth]{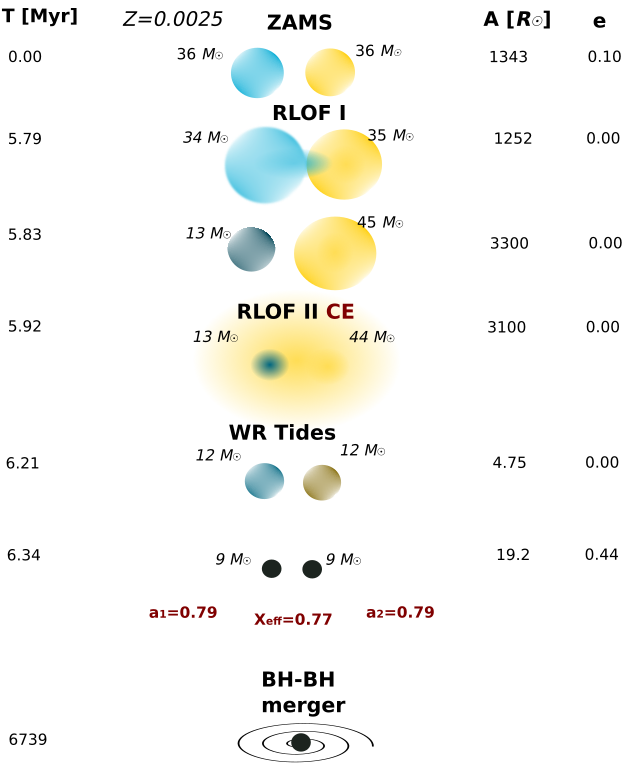}
\caption{
Typical example of evolutionary scenario with CE phase leading to the formation of BH-BH merger with 
$a_1=0.79$, $a_2=0.79$ and $\chi_{\rm eff}=0.77$. First, the binary system goes through stable 
RLOF phase with episodes of nuclear and thermal timescale mass transfer initiated by the initially 
more massive star. Then initially less massive star expands and initiates CE, after which the 
orbital separation is significantly decreased. After CE, binary hosts two compact WR stars that 
are subject to tidal spin-up. Both stars explode as supernovae and form BHs on 
eccentric orbit with merger time of $\sim 6.7$ Gyr (see Sec.~\ref{sec.res}).
}
\label{fig.evol2}
\end{figure}

Figure~\ref{fig.evol2} shows a typical CE evolution scenario (standard {\tt StarTrack} RLOF stability criteria) 
leading to the formation a BH-BH merger with both BHs spun-up by tidal interactions. At the beginning, 
the binary system of two $\sim 36 \msun$ stars with $Z=0.0025$ is on a wide ($A\approx1340\rsun$) 
and eccentric orbit ($e=0.1$). When the initially more massive star expands the system goes through a 
stable RLOF, after which the donor looses its H-rich envelope and the orbit circularizes. Soon after RLOF I, 
the system goes through another (unstable) RLOF initiated by the initially less massive companion star. 
The ensuing CE evolution leads to significant orbital contraction from $A=3100\rsun$ to $A=4.5\rsun$ and 
leaves two WR stars subject to strong tidal interactions. Both stars end their evolution at a similar time 
forming via supernovae explosions two $\sim 9\msun$ BHs. At the formation, both BHs get significant natal kicks that makes the system orbit larger 
$A\approx19\rsun$ and eccentric $e=0.44$, leading to a merger time of $\sim 6.7$ Gyr.

In Table 2 we present the statistical spin properties of BH-BH systems merging at redshifts $z<1$ for 
the two tested RLOF stability criteria models. In the rows $1-6$ we list the percentage of the BH-BH 
mergers with effective--spin parameter values $\chi_{\rm eff}>0.0,\ 0.1,\ 0.2,\ 0.3,\ 0.4,\ 0.5$. In 
the rows $7-9$ we list the percentages of BH-BH mergers with a highly spinning primary BH $a_1>0.5,\ 0.7,\ 0.9$ 
while the rows $10-12$ give the percentages of mergers with a highly spinning secondary BH 
$a_2>0.5,\ 0.7,\ 0.9$. The full distribution of the primary--spin, the secondary--spin and the effective--spin parameter for both the CE and non-CE evolution, is plotted in Figure \ref{fig.spins_distribution} in APPENDIX A.

\begin{table}
\caption{Predictions for BH-BH mergers from binary evolution}
\begin{tabular}{rccc}
\hline\hline
No. & condition$^{a}$ & CE model & non-CE model\\
\hline\hline
 1 & $\chi_{\rm eff}>0.0$  & 97\%   & 93\%  \\
 2 & $\chi_{\rm eff}>0.1$  & 95\%   & 85\%  \\
 3 & $\chi_{\rm eff}>0.2$  & 70\%   & 60\%  \\
 4 & $\chi_{\rm eff}>0.3$  & 36\%   & 39\%  \\
 5 & $\chi_{\rm eff}>0.4$  & 10\%   & 21\%  \\
 6 & $\chi_{\rm eff}>0.5$  &  2\%   &  7\%  \\
   &                         &        &       \\
 7 & $a_1>0.5$               &  3\%   &  34\%  \\
 8 & $a_1>0.7$               &  2\%   &  15\%  \\
 9 & $a_1>0.9$               &  1\%   &   1\%  \\
   &                         &        &       \\
10 & $a_2>0.5$               & 52\%   &  11\%  \\
11 & $a_2>0.7$               & 33\%   &   7\%  \\
12 & $a_2>0.9$               & 12\%   &   2\%  \\
\hline
\hline
\end{tabular}
\\$^{a}$: We list fractions of BH-BH mergers (redshift $z<1$) produced in
our two population synthesis models satisfying a given condition.\\
\label{tab.spins}
\end{table}

\section{Discussion and Conclusions}
\label{sec:concl}

The rapidly increasing number of detected BH-BH mergers does allow for some general population statements
~\citep{Roulet2021,Galaudage2021,Abbott2021b}. It appears that {\em (i)} majority ($\sim 70-90\%$) of 
BH-BH mergers have low effective spins consistent with $\chi_{\rm eff}\approx0$ and that {\em (ii)} 
small fraction ($\sim 10-30\%$) of mergers have positive non-zero spins that can be as high as 
$\chi_{\rm eff} \gtrsim 0.5$. Additionally, the population is consistent with {\em (iii)} no systems 
having negative effective spins and {\em (iv)} a not isotropic distribution of effective spins (which could indicate dynamical origin). Finally, {\em (v)} there is at least one case of a primary BH (more 
massive) in a BH-BH merger with very high spin ($a_1>0.7$ at 90\% credibility). These properties are noted to be 
broadly consistent with BH-BH mergers being formed in an isolated binary evolution. 

In our study we have tested whether we can reproduce the above spin characteristics with our binary 
evolution models that employ efficient AM transport in massive stars and that impose 
tidal spin-up of compact massive Wolf-Rayet stars in close binaries. The two presented models 
employ our standard input physics but allow for the formation of BH-BH mergers assisted either by a CE 
or by a stable RLOF. We find that the observed population and its spin characteristics ({\em i--v}) is 
consistent with our isolated--binary--evolution predictions (see Tab.~\ref{tab.spins}). In particular, 
we find that the majority of BH-BH mergers have small positive effective spins: $\sim 70\%$ mergers have  
$0<\chi_{\rm eff}<0.3$ (efficient AM transport), while a small fraction have significant 
spins: $36-39\%$ mergers have $\chi_{\rm eff}>0.3$ and $2-7\%$ mergers have $\chi_{\rm eff}>0.5$ 
(tidal spin-up). The fraction of systems with negative effective spins is small ($3-7\%$) as most BHs do 
not receive strong natal kicks in our simulations. Individual BH spins can reach high values. A large 
fraction ($11-52\%$) of secondary BHs may have significant spin values ($a_2>0.5$) as it is the less massive 
stars that are most often subject to tidal spin-up. Nevertheless, primary BHs may also form with high 
spins ($3-34\%$ with $a_1>0.5$) if both stars have similar masses and both are subject to tidal spin-up 
(see Fig.~\ref{fig.evol2}) or due to mass ratio reversal caused by the RLOF (see Fig.~\ref{fig.evol1}). We 
also note the formation of a small fraction of almost maximally spinning BHs: $2-12\%$ for $a_2>0.9$ 
(secondary BH) and $1\%$ for $a_1>0.9$ (primary BH). These results on effective spins and individual 
BH spins are consistent with the current LIGO/Virgo population of BH-BH mergers. Note that \cite{Qin2021} came to different conclusions, finding the high-spinning detections challenging for the Tayler-Spruit dynamo, especially for the unequal mass event with a high spinning primary (GW190403). Our non-CE model reproduces this type of mergers due to the mass ratio reversal (see Fig. \ref{fig.evol1}). In this channel, at the onset of the second stable RLOF, the donor may be even 5-6 times more massive than the accretor, ending as an unequal mass $(q \leq 0.4)$ BH-BH merger. \cite{Qin2021} have not considered the case of a stable RLOF in such unequal mass systems.

The above fractions correspond to just two different modes of spinning-up during the classical isolated binary BH-BH 
formation. Had we varied several other factors that influence BH spins and their orientations in BH-BH 
mergers, the ranges of these fractions would have broadened. Some obvious physical processes that can affect BH 
spins and their orientations include: initial star spin alignment (or lack thereof) with the binary 
AM, the alignment of stellar spins (or lack thereof) during RLOF phases, the treatment of accretion, the initial mass ratio 
distribution that can alter the ratio of systems going through stable and unstable (CE) RLOF,  and the natal 
kicks that can misalign spin orientations. Above all, the three major uncertainties include 
the initial stellar rotation of stars forming BHs, the efficiency of AM transport and the strength of tides in close binary systems. All of the above are only weakly constrained. Note that this is a proof-of-principle study that is limited only to BH spins in BH-BH mergers. In particular, we did not try to match BH masses and BH-BH merger rates for the highly spinning LIGO/Virgo BHs. In this study we have only shown that it is possible to produce highly spinning BHs by tidal interactions of stars in close binaries in evolution that includes and does not include CE. Our two examples of evolution (Fig. \ref{fig.evol1} and \ref{fig.evol2}) have much smaller masses than the LIGO/Virgo mergers with highly spinning BHs (Tab. \ref{tab.obs}). Note, however, that we have not used here the input physics that allows for the formation of BHs with mass over $50\msun$. Such model is already incorporated and tested within our population synthesis code~\citep{Belczynski2020c}. An attempt to match all observed parameters simultaneously is projected to happen in the future when LIGO/Virgo will deliver a larger sample of highly spinning BHs.

Given the results presented in this study, alas limited only to BH spins, we conclude that
{\em (i)} the isolated binary evolution channel reproduces well the BH spins of the LIGO/Virgo mergers {\em (ii)} if, in fact, the binary channel is producing the majority of the LIGO/Virgo BH-BH mergers, then this indicates that the AM transport is efficient in massive stars and the tidal interactions in close binaries are strong.

\acknowledgements
We thank the anonymous reviewer, Jean-Pierre Lasota, Ilya Mandel and Sambaran Banerjee for their useful comments on the manuscript.  
KB and AO acknowledge support from the Polish National Science Center (NCN) grant
Maestro (2018/30/A/ST9/00050).

\bibliography{biblio}

\section*{APPENDIX A}

\begin{figure*}[h]
\hspace*{-0.4cm}
\includegraphics[width=0.85\textwidth]{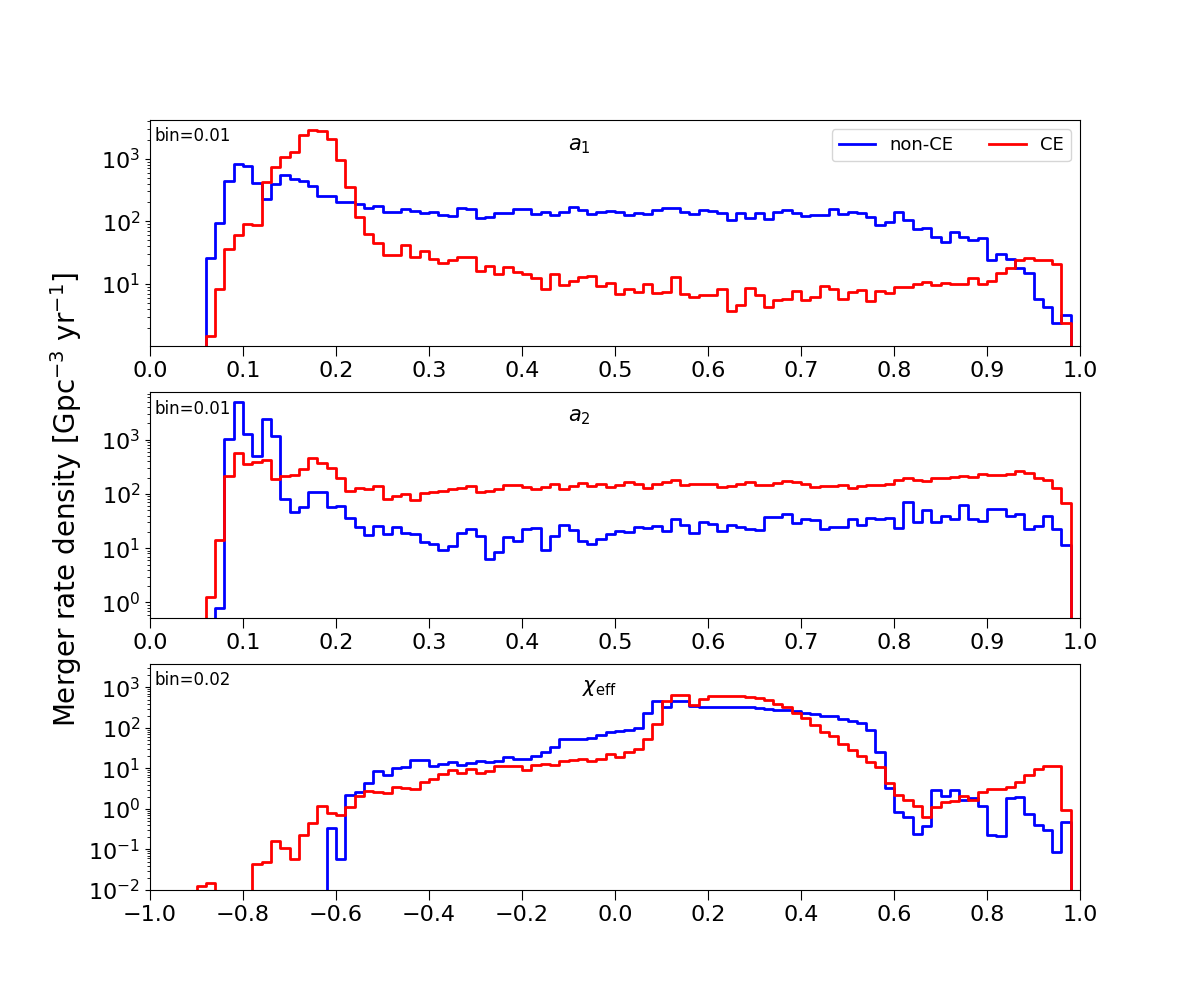}
\caption{
{Distribution of primary BH spin ($a_1$) -- top panel; secondary BH spin ($a_2$) -- middle panel; effective spin parameter ($\chi_{\rm eff}$) -- bottom panel; of BH-BH mergers at redshifts $z<1.0$. The results are for two tested models: the non-CE model plotted with red line and the CE model plotted with blue line. The figure is a supplement to the statistical spin predictions shown in Table \ref{tab.spins} and described in Section \ref{sec.res}.}
}
\label{fig.spins_distribution}
\end{figure*}

\end{document}